# Quantum Printing[TM]

G. Aeppli[1,2,3], A Balatsky[4,5] , S. Bonetti[6,7],  G. Cardoso[5], S. Raghu[8], E. Syljuåsen[9], T.-T. Yeh[4], S.-Z. Lin[10,11], Y. Liu[5], J. Weisenrieder[12], P. Wong [4,5]

[1]Paul Scherrer Institut, Villigen, PSI CH-5232, Switzerland [2]Institut de Physique, EPFL, Lausanne CH-1015, Switzerland, [3]Department of Physics and Quantum Center, ETH Zürich, Zürich, CH-8093, Switzerland; [4]Department of Physics, IMS, University of Connecticut, Storrs, Connecticut 06269, USA; [5]Nordita, Stockholm University, and KTH Royal Institute of Technology, Hannes Alfv´ ens vag 12, SE-106 91 Stockholm, Sweden; [6]Department of Molecular Sciences and Nanosystems, Ca' Foscari University of Venice, Venice, Italy; [7]Rara Foundation – Sustainable Materials and Technologies, Venice, Italy; [8]Department of Physics, Stanford University, Stanford, CA, USA; [9]Center for Quantum Spintronics, Department of Physics, Norwegian University of Science and Technology, NO-7491 Trondheim, Norway;[10] Theory Division and [11]Center for Integrated Nanotechnology, Los Alamos National Laboratory, Los Alamos, New Mexico 87545, USA;  [12]Light and Matter Physics, School of Engineering Sciences, KTH Royal Institute of Technology , SE-100 44 Stockholm, Sweden

Abstract:

We introduce the concept of *quantum printing*—the imprinting of quantum states from photons and phonons onto quantum matter. The discussion is focusing on charged fluids (metals, superconductors, Hall fluids) and neutral systems (magnets, excitons). We demonstrate how structured light can generate topological excitations, including vortices in superconductors and skyrmions in magnets. We also discuss how quantum printing induces magnetization in quantum paraelectrics and strain-mediated magnetization in Dirac materials. Finally, we propose future applications, such as printing entangled photon states, creating entangled topological excitations, and discuss applications of quantum printing to light induced quantum turbulence in a charged fluid. This review represents the expanded version of the shorter review submitted to Nature Physics.

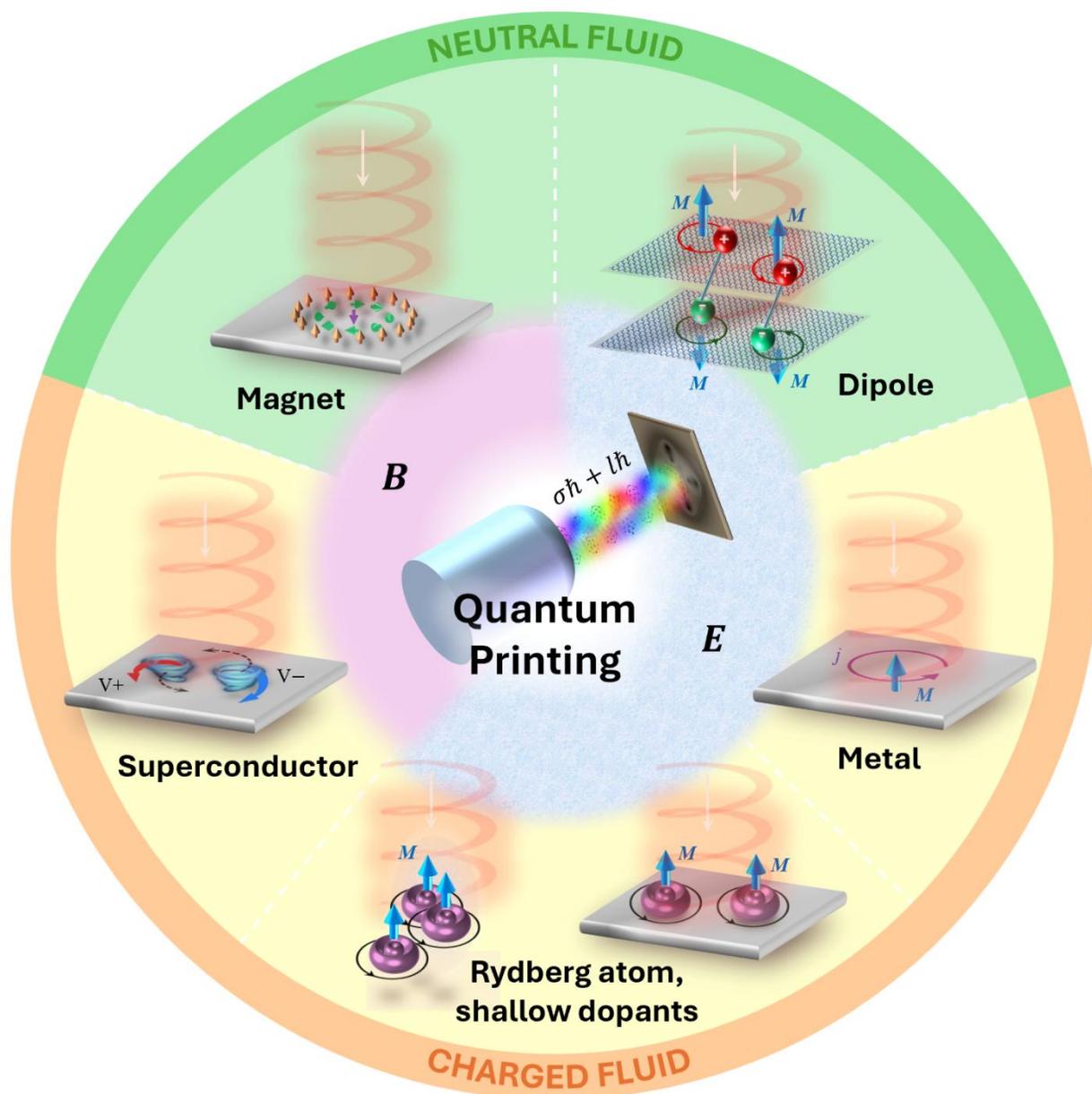

**Fig. 1:** Schematic diagram illustrating the applications of quantum printing in neutral fluids (green area) and charged fluids (red area). Neutral fluids include examples such as magnetic and dipolar systems, while charged fluids encompass metals, trapped ions, superconductors, like shallow dopants and Rydberg states and related materials. The quantum printing effect can involve structural imprinting driven by the magnetic $\boldsymbol{B}$ or electric $\boldsymbol{E}$ fields of light. Similar QP can be accomplished with strain and phonons.

***Content***

***Notations***

We are listing often used notations in the paper.

$M$ = magnetization

$A$ = gauge potential

$E$ – electric field

$B$ - magnetic field

$E_\omega$ - AC field

$e_{x,y}$ – polarization axes of the field

$\varphi_{az}$ – azimuthal angle of electric field of the LG beam in the plane perpendicular to propagation axis

$\sigma_\omega^L$ - AC longitudinal conductivity

$\sigma_\omega^H$ - AC Hall conductivity

$\xi_0$ - superconducting coherent length

$\mu_B = \frac{q\hbar}{(2m_*)}$ - Bohr magneton

LP – linearly polarized light

CP – circularly polarized light

SAM- spin angular momentum

OAM- orbital angular momentum

RO- radial order

QP- quantum printing

***Introduction***:

Recent advances in ultrafast light-matter experiments have enabled unprecedented control over quantum systems culminating in the dynamical induction of novel states of quantum matter by light. The concept of engineering quantum matter with desired properties on demand has consequently emerged. This approach naturally requires the active manipulation and "steering" of quantum matter to the desired state. Pioneering work on energy transduction and collective mode excitations has laid the groundwork for manipulating quantum phases in the time domain using light [Bas2025,Nov2017,Ker2025] and strain [For2011,Du2023,Hor2020a].

New opportunities emerge in this field where we could take advantage of structured light and strain sources with inherent spin and orbital angular momentum structures. We now can transfer photon and phonon angular and spin quantum numbers to matter and stir it to desired states. To illustrate the role of structured beams impinging on quantum matter we bring the concept of **quantum printing (QP)**, see Fig.1 where structured light—carrying quantized angular momentum, spin, or chirality—directly imprints quantum numbers into matter, inducing tailored orders (e.g., magnetization, chirality, vortices) or rectified nonlinear response. For an explanation of structured light, see Fig. 5. Ideas of transduction of quantum numbers have been experimentally verified in the Inverse Faraday Effect where circularly polarized light induces dc magnetization in quantum paraelectrics [Bas2024], ultrafast Barnett effect in magnets (lattice to magnetization of chiral phonons [Dav2024], interplay with magnetization)[Rom2024] and photo-induced chirality [Zhe2025].

This review outlines some of the current work and proposes a few future directions for applications of QP and provides the timely discussion for the rapidly growing field and a possible roadmap for future research. We start with the brief history overview of QP and point potential for use of QP with entangled states and the relation to the no-cloning theorem. We then consider in more detail the cases of A) charged quantum fluids i.e. the electron and superconducting fluids. We start by outlining some of the ideas on classical inverse Faraday effect as an earlier example of QP with circularly polarized light and discuss the effects of dynamic induction of magnetization via dynamical multiferroicity [Bas2024]. Furthermore, among coherent electron states we consider the case of A1) superconductors, A2) Rydberg and semiconductor dopant states A3) Acoustic induction of magnetization, A4) quantum Hall fluids, i.e. the transverse fluids where there is a significant off-diagonal conductivity response, A5) the outlook for light induced quantum turbulence and B) the newer and less explored case of neutral fluids, such as B1) particle- hole spatially separated excitons in bilayer materials, B2) the case of quantum printing of topological excitations in magnets. The distinction between charged and dipolar fluids is not sharp. For example, Rydberg states and dopant states on semiconductors can be equally viewed as dipolar states. The distinction we like to use is the electrons in these cases are moving under light while the ion and dopant atoms are assumed to be fixed. In that regard the analysis of response of these states is very similar to a response of electrons in superconductors. In a broader view even electrons in metals and in superconductor are dipolar states: they propagate in the presence of neutralizing positive ions whose dynamics is ignored. This is not the case for the excitonic states where both positive and negative charges are moved by light. It is this focus on one component response is how we delineate the case of charged vs neutral fluid.

***History***. The idea that light and strain affect matter is central to materials science and is fundamental to technology. Ideas of printing permeate our culture and technology, starting most notably with photography in the nineteenth century, whose principles underpin today's photolithography which is key for the production of the integrated circuits upon which the information revolution depends. We are talking about the step beyond photography: crucial feature of classical image creation in photography is that it leaves behind a classical record, and so represents classical printing, although the underlying photochemistry is intrinsically quantum mechanical. In our view, quantum printing really starts with Rabi, who pointed out that light can induce the now-called 'Rabi' oscillations of the level occupancies of quantum systems [Rab1936]. Such oscillations have no analog in the classical world, where descriptions of pumped matter rely simply on population dynamics and interference effects do not play a role. The combination of the underlying concept of the Bloch sphere and entanglement have given rise to magnetic resonance [Abr1992], quantum optics and now the general fields of "quantum technologies" where photons of various wavelengths are used to "write" and "read" quantum states, primarily for real and artificial atoms and ions.

Structured light QP was also discussed in the 30s. The transfer of angular momentum (AM) from light to matter dates back to the pioneering work of Bethe in the 1930s, where the SAM of light was first confirmed through a mechanical torque measurement on a suspended wave plate [Bet1936]. This foundational result provides the basis for the modern concept of QP—the transfer of optical quantum number and rotational properties such as SAM and OAM to microscopic and macroscopic systems. In atomic-scale systems, QP has been realized in atoms, ions, and molecules, where the discrete nature of light's quantum numbers couples with the quantized energy structure of matter. Notable examples include spin–orbit excitations of Bloch electrons [Mas2024], molecular transitions [Mon2014], Rydberg ions [Muk2018].

Techniques such as electromagnetically induced transparency (EIT) [Aki2015] have further enabled light-matter imprinting with high coherence. Quantum printing has even been implicated in high-energy phenomena such as positron pair production [Zhu2018a]. At mesoscopic scales, optical angular momentum has been harnessed to rotate nanoparticles [Fu2019,Leh2013,Sim2010,He1995], controlled exchange of internal and external AM to Bose–Einstein condensates (BEC) [Che2018,Wri2008], and induce currents in 2D materials [Nor2024, Kar2010], etc. On macroscopic scales, quantum printing emerges in condensed matter through control of collective excitations, such as magnetization [Sta2007], Higgs modes [Miz2023,Kan2025], or spin-polarized photocurrents observed via the circular photogalvanic effect, topological insulators and Weyl semimetals [Jua2017, Tu2017, Mcl2012, Gan2002], circular or chiral phonon magnetization [Bas2024,Zhu2018b]. Besides, in analogy to thermo-printing via heating paper, quantum printing also allowed the microkelvin atomic cooling via AM transfer to the $\sigma+\sigma-$ process [Coh1990]. Classical printing assumes passive response of the paper on which the paper retains the effect of the beam. Quantum printing opens the opportunities of backaction where the quantum matter induces new modalities of light interacting with the medium. Recent advances even suggest a reverse process, where the magnetic texture of materials patterns photon fields via magneto-optic effects [Kim2022], offering a compelling reciprocity to the concept of quantum printing: light is modified by the quantum state of matter. We also mention the first proposal about the generation of OAM carrying beam by Allen et al [All1992].

***Quantum printing (QP).*** We start with classical printing where the matter (ink) was transferred from original to copy (clean parchment) [Lu2023,Tsi1985,Wol1975,Sab2000]. The laser printing invented in the 70s in Xerox Palo Alto was a first example of printing with light where the ponderomotive effects of light affected the surface and induced the readable character with no ink being transferred [Cad2020]. Now, simple heating due to light allows printing (recall your "heat printed "credit card receipts). The next step in this evolution is the notion of quantum printing where the quantum numbers of light (orbital (OAM) and spin angular momentum (SAM)) are transferred to the quantum material. The phase coherence of the quantum state can lead to qualitatively different behavior of quantum matter subject to structured light interaction, Fig 2,3.

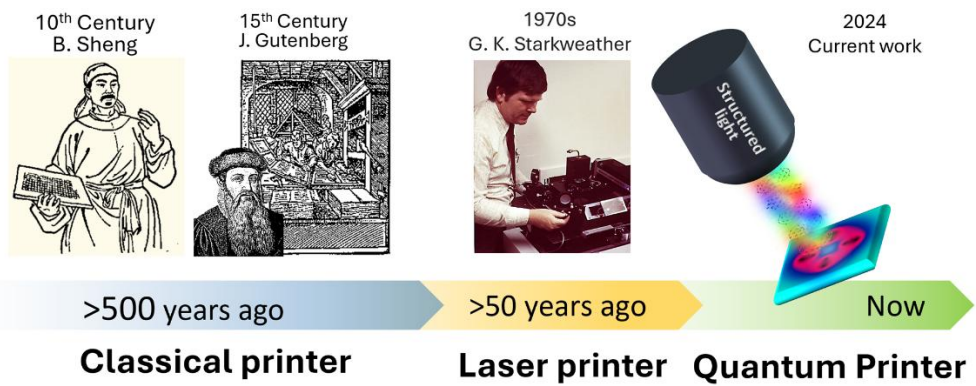

**Fig 2:** Illustration of the concept of quantum printing as a natural evolution of printing concepts [Hai2023, Wol1974, Wag2000, Met2020].

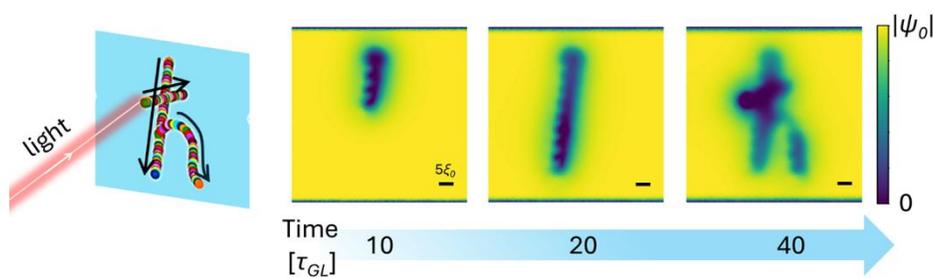

**Fig 3:** to further illustrate the concept we demonstrate the quantum printing of the Planck's constant $\hbar$ on the superconducting surface with circularly polarized light. The phase and amplitude of SC dynamics is shown in the linked video [SM]. The quantum nature of the printing in our case is supported by sensitivity of the amplitude and phase changes of coherent electron fluid to photon quantum numbers and by induction of the quantum topological excitations- vortices, as we illustrate below. The scalar bars in the snapshots represent 5 superconducting coherent lengths (Movie S1 in Supplemental Material [SM]).

We start with an early example of *QP*: the Inverse Faraday Effect (IFE) – the induction of the dc magnetization in a material illuminated by circularly polarized light $E_\omega = E_0 \left( e_x + i e_y \right) e^{i\omega t}$ (case of σ = π/2 in SAM of Fig. 5), Fig.4 . The chirality of the light allows the dc and time-odd symmetry combination $\boldsymbol{M} \sim \boldsymbol{E}_\omega \times \boldsymbol{E}_{-\omega} \sim \boldsymbol{E} \times \partial_t \boldsymbol{E}$ that predicts the dc magnetization induced [Her2006,Kir2010,Van1965,Sha2025]. The second form of IFE expression is less familiar but equally valid. The established and tested IFE in our context proves that net magnetization can be printed in nonmagnetic media with chiral light [Rei2010] or other chiral fields – phonons [Bas2024].

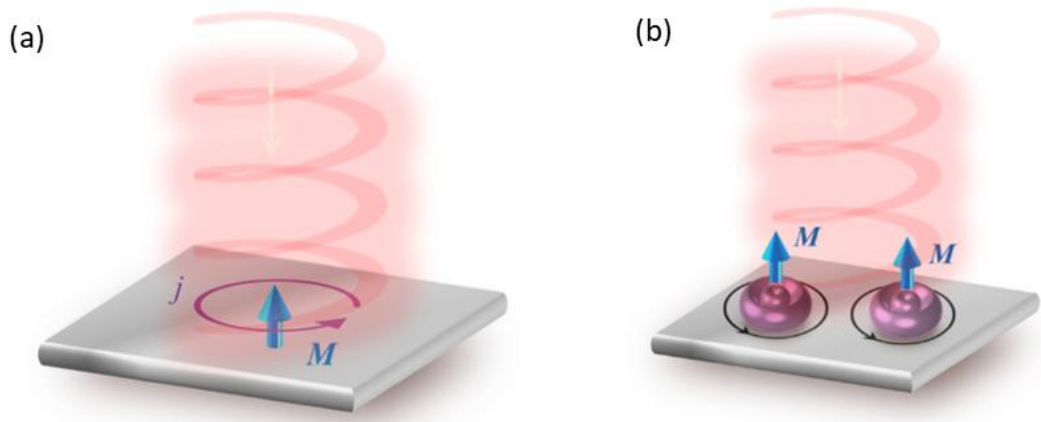

**Fig 4.** Schematic of light-induced magnetization and charge currents via the IFE in (a) metals and (b) in shallow dopants in semiconductors. (a) Orbital magnetic moment of electron flow results in the magnetic moment of charged flow. The magnetization induced can be expressed as a magnetic moment of electrons in an effective external field. No external static field is applied in IFE. Magnetization is generated via orbital motion. (b) the IFE induced orbital magnetization of the shallow dopant and Rydberg state. See the discussions in A1.2 and A2.

This example contains some general features we find across multiple examples of quantum printing: i) the transient nature of the induced imprinted state- the state like dc magnetization persists as long as imprinting is going on. Once the field is switched off the state will decay, with lifetime limited by T1 and T2 processes. While the state decays it will possess the "imprinted" properties like magnetization in the above example. One can indeed assume nontrivial response of the quantum matter and hence what we discuss here can be viewed as an example of active quantum matter. The driven quantum matter develops nontrivial transient orders that are not present at equilibrium; ii) the states that are particularly susceptible to quantum printing ("paper") are quantum coherent states where charge, lattice and spin coherence enables a good imprinting of photon and phonon quantum numbers; iii) the ability to print excitations with light implies reciprocal ability to erase. For example, one can use the QP to reverse spin magnetization and erase vortices.

***Laguerre–Gaussian beam***.  Laguerre–Gaussian (LG) beams are widely studied structured light beams that carry orbital angular momentum (OAM), characterized by a helical phase structure $e^{-il\varphi_{az}}$, where $l$ determines the OAM $l\hbar$ per photon and $\varphi_{az}$ is the azimuthal angle. These beams, first formally described by Allen et al. in the early 1990s [All1992], are solutions to the paraxial wave equation in cylindrical coordinates and are defined by two mode indices: the azimuthal mode index $l$, and the radial order (RO) $p$.  For nonzero OAM LG beams exhibit intensity node in the center of the beam and a characteristic doughnut-shaped intensity distribution. They can also carry spin angular momentum (SAM) $\sigma\hbar$ through polarization modulation.

Fig5 illustrates the time evolution and intensity profiles of selected LG modes using the following electric field $E_{LG}(r,t)$:

$$E_{LG}(r,t) = u_{p,l}(r,\varphi_{az})\left(\cos\theta_{pol}\ x + e^{i\sigma\frac{\pi}{2}}\sin\theta_{pol}\ y\right),$$
$$u_{p,l}(r,\varphi_{az}) \propto E_0(r)\ L_p^l e^{-il\varphi_{az}},$$

where $u_{p,l}(r,\varphi_{az})$ is the LG transverse mode expressed with the term Gaussian beam $E_0(r)$, $L_p^l$ represents the Laguerre polynomials.

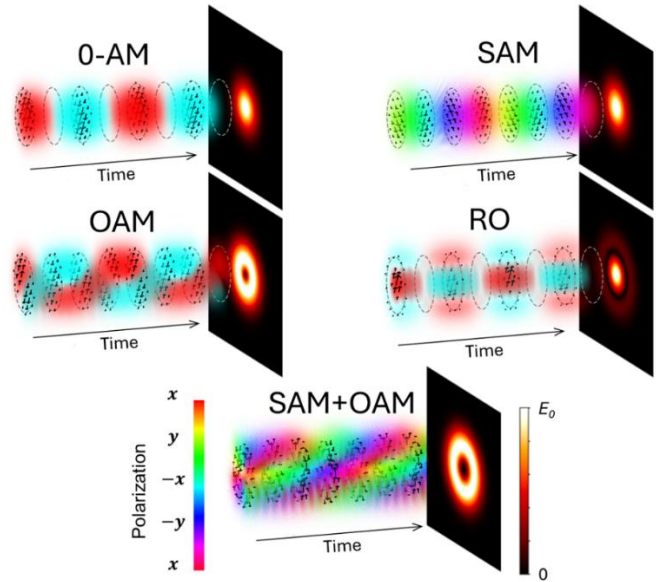

**Fig 5:** Time evolution and intensity profiles of representative LG modes $LG_{pl}$. Demonstrated modes include 0 -AM: Linearly (LP) polarized $LG_{00}$, SAM: Circularly polarized (CP) $LG_{00}$, OAM: LP $LG_{01}$, RO: LP $LG_{10}$, and combining multi-effects SAM+OAM: CP $LG_{02}$. The time evolution of LG modes is visualized through changes in intensity and polarization, where the polarization direction is represented by color hue, and the field amplitude is indicated by the color density. Dashed circles indicate beam

spot positions at time steps $\omega t$ =0, π/2, π, ..., 7π/2, and the polarization projected on the transversed slice as arrows. The right side of each time trace shows the amplitude projection of the electric field. Color bars at the bottom represent the polarization state and electric field amplitude. LG beams are typically generated using spatial light modulators (SLMs), spiral phase plates, or q-plates [Wan2012,Mar2006, Bei1994]. They are prototypical examples of light carrying OAM, commonly referred to as twisted light or vortex beams, and have become foundational tools in optical angular momentum research.

**QP, entangled states and no-cloning.**

QP has important connections with quantum information concepts. We outline the questions we see emerging in this context:

*QP and entangled states*. Most of the QP applications we address would be possible with the coherent macroscopic beams. The extension of the QP to the case of entangled photons is a fascinating topic to be developed. Consider the LG beam with $|l = 2\rangle$ that can be downconverter to two $|l = 1\rangle_{A,B}$ photons that are entangled and are impinging the printed surface in different spots labelled A and B. One can imagine that we can print two collapsed states of the former *l*=2 beam $|l = 2\rangle$: $|l = 2\rangle \rightarrow |l = 1\rangle_A \otimes |l = 1\rangle_B$ . The distribution of intensisies and phases would encode the entangled nature of the entanglement of two photons.

*QP and no-cloning*. The question can be asked about connection and constraints on QP arising from the no-cloning theorem for quantum states. There is no tension between QP and no-cloning. No cloning theorem prohibits unitary operation creating the exact copy of the initial state [WIK2025]. QP allows for initial state to be destroyed via conversion of photon/phonon into collective excitations in quantum matter. QP on the contrary relies on the interaction of the incoming state with the quantum matter to record it. The photon state collapses upon QP, which in its simplest form is either an absorptive or Raman event leaving no or one photon behind.

**Excitations induced by spatially structured light and strain.**

Majority of applications we discuss deal with photons. Phonons and engineered strain can be also used for QP, albeit the list of examples we can point to is smaller. QP with photons can trigger a coherent lattice response by launching phonons, which in turn modulate the material's electronic structure. A compelling example is a study on FeSe, where a combination of time- and angle-resolved photoemission spectroscopy (trARPES) and time-resolved X-ray diffraction revealed a pronounced enhancement of electron–phonon coupling, mediated by underlying electron–electron interactions [Ger2017]. Or in studies of the influence of optically driven shear modes (of A1 symmetry) on the lattice and electronic properties of Td-WTe$_2$—a type-II Weyl semimetal—demonstrating that excitation of interlayer shear modes can transiently and reversibly drive the system in and out of a topologically trivial state [Sie2019,Ji2021]. Furthermore, static strain has been shown to open transition pathways to metastable phases that are otherwise inaccessible through optical excitation alone [Ji2023,Wu2024].

QP allows the control beyond the temporal domain to also encompass spatial manipulation, which requires tools that operate at the intersection of femtosecond temporal resolution and nanometer-scale spatial precision, Fig 6. Spatially structured light—optical fields with engineered intensity and phase profiles—offers a promising route to realize such control [Vas2024]. Deliberate interference of laser beams can be employed to form transient optical gratings (TOGs) or lattices (TOLs), periodic excitation patterns with tunable symmetry, adjustable periodicity, and nanometer-scale modulation of the absorbed fluence. These optical structures confine light–matter interactions with exceptional spatial and temporal resolution, serving as dynamic templates for selective phonon excitation, strain wave control, and spatially resolved modulation of the potential energy landscape in crystalline solids—capabilities that extend beyond what is possible with uniform excitation. TOG has also been successfully applied in studies of ultrafast dynamics of magnetic systems at the nanoscale. [Cao2021,Fan2025,Kse2021]

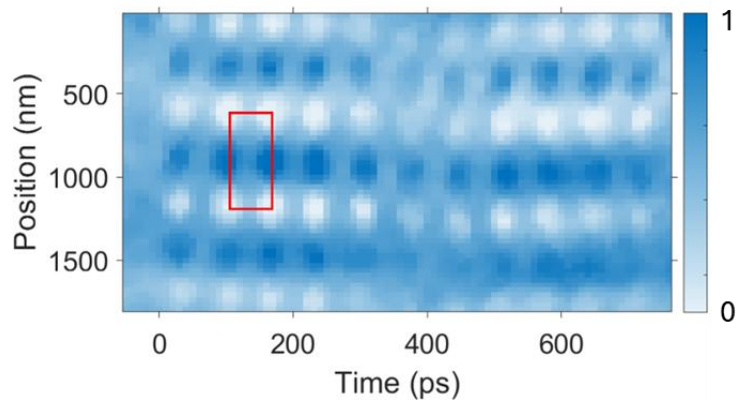

**Fig 6.** Figure shows an example of phonon QP where one can induce transient phonon cavities in a solid. A space time contour plot of structural oscillations in CrSBr excited by a TOG. The red rectangles indicate a phonon cavities in the space-time domain. Reproduced from [Hor2020a].

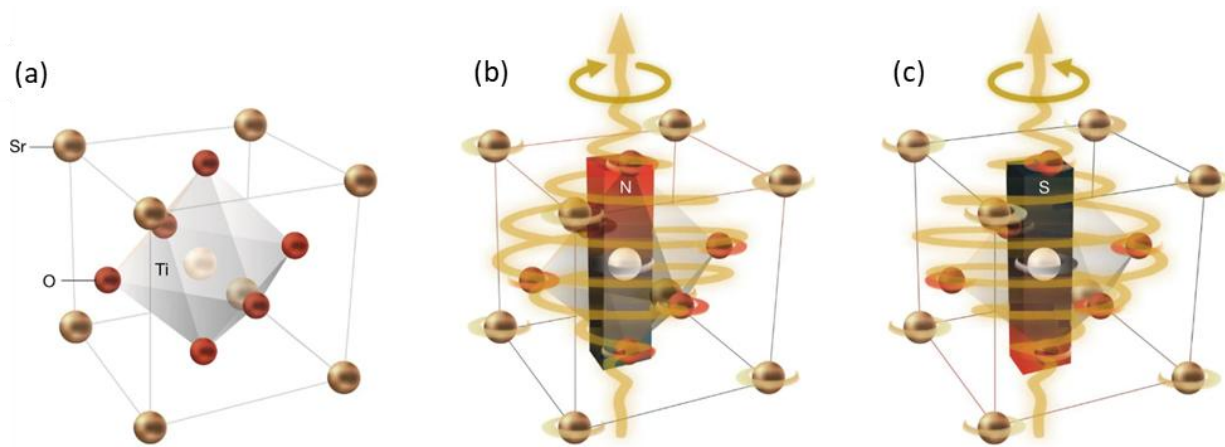

**Fig 7.** (a) SrTiO3 unit cell in the absence of a terahertz electric field. When a circularly polarized terahertz field pulse drives a circular atomic motion, dynamical multiferroicity is expected to create a net magnetic moment in the unit cell. (b),(c), The north pole points up for a pulse that is left-handed (b), and down for a pulse that is right-handed (c). [reprinted from [Bas2024]]

### QP Induced Chirality and Symmetry Breaking.

Recent breakthrough demonstrated the concept of dynamical multiferroicity in paraelectric SrTiO$_3$ (STO), showing that it is possible to imprint a magnetic moment using intense circularly polarized terahertz (THz) electric fields [Bas2024]. By resonantly driving the soft infrared-active phonon mode around 3 THz, the ions in STO undergo coherent circular motion within the lattice. This circular ionic trajectory effectively generates a net magnetic moment along the axis of rotation, analogous to a phononic version of the Einstein–de Haas or Barnett effects, Fig 7. Phononic Barnett film has also been demonstrated in a thin ferrimagnetic film [Dav2024].

In this approach, the handedness of the THz field determines the direction of the induced magnetic moment: left-handed circular polarization leads to one magnetic polarity, while right-handed polarization reverses it. This helicity-dependent magnetization is probed via time-resolved magneto-optical Kerr effect measurements, revealing clear signatures at low (~0.6 THz) and high (~6 THz) frequencies. The amplitude of the magnetic signal scales quadratically with the applied electric field, consistent with a dependence of $M \propto |E|^2$.

A theoretical description using coupled nonlinear phonon oscillators and *ab initio* self-consistent phonon calculations reproduces these observations, highlighting the critical role of the phononic Barnett effect, which transfers angular momentum from the lattice vibrations to the electronic system. The magnitude of induced magnetization is the subject of ongoing debate. This mechanism further establishes a new pathway for ultrafast, purely electric-field-driven magnetic transition, offering perspectives for THz-speed magnetization switching and novel functional devices leveraging lattice dynamics.

We now proceed with the description of specific examples where QP could be further explored. We structure this review along the two major lines: QP in charged fluids (metals, superconductors, Dirac/Weyl semimetals and QP in neutral fluids like excitons and magnets with no free carriers. The distinction is ambiguous, as the case of ferroelectric state can be equally considered in neutral section. Nevertheless, the proposed structure helps to organize the narrative.

### A. Quantum printing in charged fluids

### A1. Quantum Printing in superconductors

**i)** *Vortex in superconductors via spin and orbital angular momentum (SAM+OAM) photons*.

The superconducting (SC) state offers a compelling platform for quantum printing, owing to the unique properties of its complex order parameter. In particular, the phase of the order parameter can be imprinted by a vector potential, enabling the system to respond coherently and sensitively to external electromagnetic fields. This phase flexibility governs key physical phenomena such as supercurrents, quantized vortices, and light–matter coupling. Because a spatially varying phase corresponds directly to supercurrent flow and topological excitations, structured light can imprint its phase profile onto the SC fluid, inducing quantized vortices without the need for magnetic fields or mechanical rotation. As a result, superconductors can be utilized as a dynamic quantum recording medium, capable of encoding the spatial and temporal structure of light into their macroscopic coherent state, laying the foundation for quantum printing.

To understand how supercurrents and vortices are generated by light in superconductors, we consider the phase of the order parameter influenced by the electromagnetic vector potential. Neglecting amplitude variations, the spatial structure of vector potential $A$ governs the phase, and the resulting vorticity of supercurrent links directly to the longitudinal magnetic field component $B_{\parallel}$ [Yeh2024b]

$$\nabla \times J_s = \rho_s \nabla \times \nabla \theta_s - \rho_s \nabla \times A \propto -\rho_s \nabla \times A = -\rho_s B_{\parallel} \qquad \text{A.1.1}$$

Such fields naturally arise in non-plane-wave beams with spatial variation, such as Gaussian or vortex beams, enabling light to imprint phase structure onto vorticity of supercurrents. If the light intensity exceeds the magnetic flux quantum threshold, this can lead to ultrafast, contactless generation of vortices with zero applied magnetic field.

This light-induced vortex generation has been theoretically predicted within the time-dependent Ginzburg–Landau (TDGL) framework [Yok2020,Pla2022,Ori2020,Yeh2024a, Yeh2024b], Accordingly, Yokoyama, Plastovets et al. [Yok2020,Pla2022] proposed that the angular momentum of structured light can be transferred to the superconducting condensate with the pre-quenched process. Oripov et al. and Wang et al. also explored a vortex response in rf cavity, which rely on the geometry of edge and impurity effects. [Ori2020,Wan2024] Most recently, Yeh et al. [Yeh2024a,Yeh2024b] reported a series of works demonstrating direct printing of optical quantum state into superconducting films using Gaussian and LG beams. Their results revealed coherent, light-driven dynamics of vortices under structured electromagnetic potentials, confirming the feasibility of quantum printing via THz/sub-THz light. This approach opens a path for transferring light structure to matter. These results indicate the potential for applications of QP such as topological defect control, vortex patterning, and even THz-scale quantum memory, Fig 8.

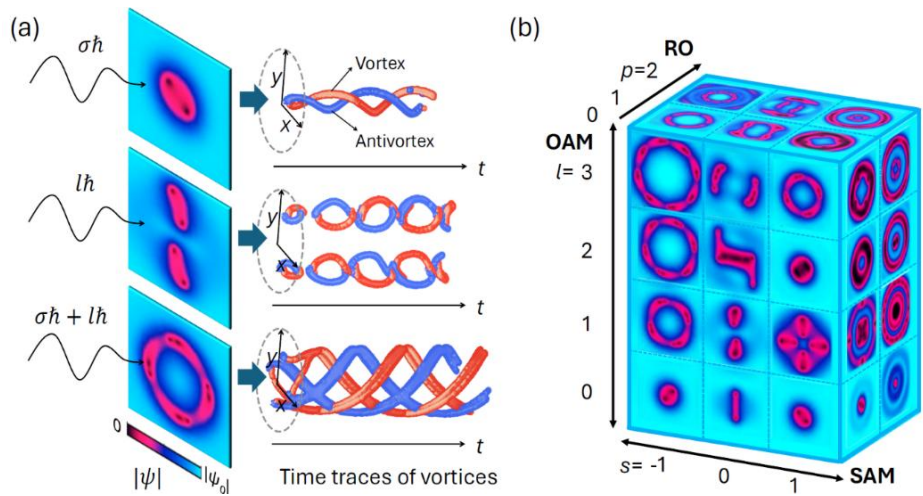

**Fig 8:** Results from quantum printing and the time evolution of light-induced vortices. (a) These results demonstrate that light carrying different quantum numbers can be coherently transferred to the superconducting state and imprinted onto vortices, as shown by the profiles of the order parameter $|\psi|$ and the corresponding time traces of the vortices. (b) Snapshots of $|\psi|$ profiles imprinted with different quantum numbers, including spin angular momentum (SAM: $s$), orbital angular momentum (OAM: $l$), and radial order (RO: $p$). (Movie S2 in Supplemental Material [SM])

### *ii)* *Metals and superconductors: Light-induced magnetization and charge currents.*

Beyond the direct printing of the optical longitudinal magnetic field onto matter, light-driven stirring of charged fluids offers another pathway for quantum printing in metals and superconductors. Circularly polarized light can transfer its SAM to the electron fluid via the nonlinear IFE, generating a static magnetization either through intrinsic magnetic responses [Lia2021,Jur2017,Qai2016,Tag2011,Ede1998] or light-induced microscopic structural changes [Sha2024,Hur2018]. Notably, this mechanism allows for ultrafast, non-contact control of magnetic order, enabling the imprinting of magnetic and electronic textures in quantum materials.

Recent microscopic theory of the inverse Faraday effect (IFE) in metals, using a Keldysh and Eilenberger quasi-classical formalism demonstrated that circularly polarized THz light can induce significant orbital magnetization purely through the orbital motion of electrons without requiring spin polarization or spin–orbit coupling [Sha2024]. Calculations predict light-induced magnetic fields up to 0.1 T in clean metals such as Nb, highlighting a robust optical imprinting mechanism in the absence of external fields, Fig.4. Hurst et al. [Hur2018] explored the IFE in plasmonic gold nanoparticles using a quantum hydrodynamic model. They showed that resonant excitation with circularly polarized light generates a static orbital magnetic moment via collective surface plasmon oscillations. The resulting magnetization driven by surface-bound rotating electron currents reaches values as high as 0.35 μB/atom under moderate laser intensity, and scales with both nanoparticle size and laser power. These studies demonstrate that both extended metallic systems and nanoscale plasmonic particles can exhibit strong, nonthermal magnetic responses to circularly polarized light, providing a foundation for optically driven quantum states in metallic and superconducting systems.

These findings point toward a new regime of quantum printing, where helical light fields inscribe microscopic magnetic textures and charge flows into conducting or superconducting states. Unlike conventional lithography or current-driven methods, this approach exploits quantum coherence and nonequilibrium dynamics to reshape matter optically, opening new possibilities for dynamically reconfigurable quantum materials and devices.

### A2. IFE in Semiconductors and Rydberg Systems

A particular example of the IFE is that of inducing the effect in Rydberg states [Won2025]. Rydberg atoms are isolated atoms with a valence electron occupying a state with $n \gg 1$, which results in a highly extended radial wave function and a large electric dipole moment. The large radial extent of the Rydberg wave function means that the outer electron experiences a very weak (shallow) potential. The shallowness of this potential allows Rydberg atoms to be highly sensitive to external electric fields which makes them an attractive platform also for metrology; switching atoms into and out of such states also provide the basis for an increasingly popular paradigm for quantum processors [Luk2001,Hen2020,Blu2024] (Fig.4 B). Rydberg atom-based platforms can be sensitive to electric fields on the order of $10^{-1}$ V m$^{-1}$. Rydberg atoms can be adopted as precision radio frequency sensors, which are highly tunable to specific frequencies and are robust to noise. Moreover, these states are highly sensitive to magnetic fields as well. The large extended electronic wave function of the Rydberg state lends itself well to generating a large orbital magnetic field induced by circularly polarized light.

The effective Hamiltonian describing the orbital magnetization induced by the IFE takes the form of a Zeeman term $H_{\text{eff}} = \mu_B \mathcal{B}_{\text{eff}}$, reading as

$$\langle a|H_{\text{eff}}|b\rangle = \mu_B \left[ \frac{2m_*q}{\hbar^2} \omega (\mathcal{E}_R^2 - \mathcal{E}_L^2) \times \sum_c \left( \frac{\langle a|r_+|c\rangle\langle c|r_-|b\rangle}{\omega_{bc}^2 - \omega^2} - \frac{\langle a|r_-|c\rangle\langle c|r_+|b\rangle}{\omega_{bc}^2 - \omega^2} \right) \right] \qquad \text{A.2.1}$$

where  is the effective Bohr magneton and the quantity in the brackets has units of magnetic field strength. We emphasize here that there is no real magnetic field present, however the atoms respond as if they possessed a unit Bohr magneton magnetic moment immersed in a magnetic field of strength $\mathcal{B}_{\text{eff}}$.

For Rydberg-like states, the states $|a\rangle$, $|b\rangle$, and $|c\rangle$ in the expression for the effective Hamiltonian matrix elements are taken to be hydrogenic states defined in terms of their principal quantum number $n$, orbital angular momentum number $\ell$, and magnetic quantum number $m$, such that the states may be labelled by the tuple $|a\rangle = |n, \ell, m\rangle$.

Rydberg states in isolated atoms produce an effective magnetic field which scales as $n^4$. With a 1THz driving beam of intensity , effective magnetic field printed is on the order of $1\mu T\ n^4$ for typical Rydberg atoms such as rubidium and cesium [Won2025].

In addition to free atoms, Rydberg states also occur for dopants in semiconductors [Gre2010,Mur2013,Ste2009,Van2018]. On account of the high dielectric constants of the semiconducting hosts, these dopant atoms also exhibit large extended electronic wave functions, even for low values of the principal quantum number, such as $n < 10$. For example, the Bohr radius for Si:P is 3.17nm, roughly two orders of magnitude larger than that for a free atom. This means that lower lying exited states can be used as well. For Si:P the induced effective magnetic fields are on the order of $10^2$T for states of $n \sim O(1)$ for a beam intensity of [Won2025]. The large magnitude of this effect demonstrates that even for seemingly simple systems, non-linear effects build up dramatic results, Fig.9.

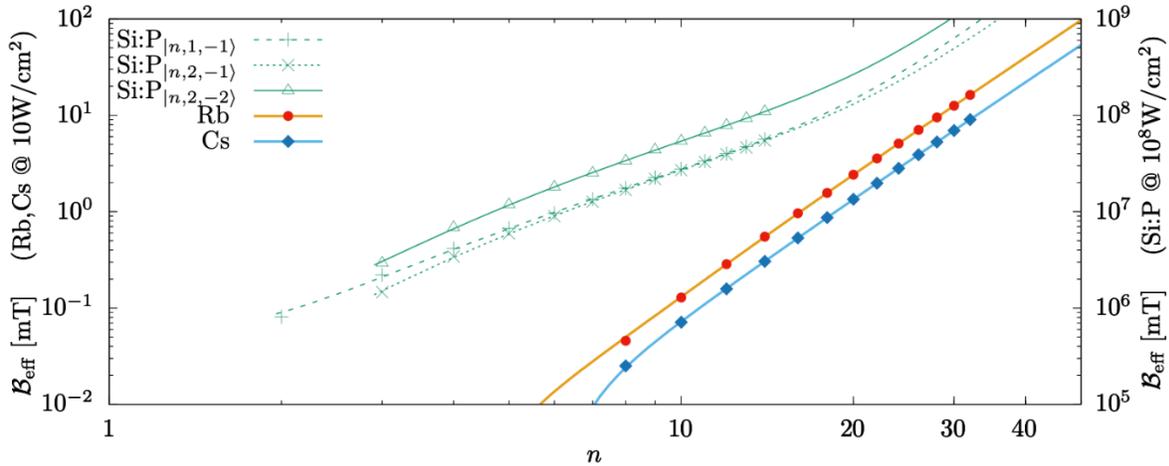

**Fig 9:** Magnitude of the IFE for rubidium and cesium Rydberg atoms, as well as various states of Si:P, for differing values of the principal quantum number $n$.

### A.3 Acoustic Inverse Faraday Effect: Acoustic Quantum Printing of Magnetization

QP using strain and phonon fields offer an equally powerful and less explored route to induce and control states [Jur2017,Jur2020,Sha2024]. We focus on examples of acoustic inverse Faraday effect (IFE) [Su2025] and axial magnetoelectric effect [Lia2021] where a circularly polarized acoustic wave induces a static orbital magnetization. These effects are especially promising in Dirac materials, where they emerge purely from the interplay between strain-induced lattice dynamics and the intrinsic band topology of the system. [Su2025]. To illustrate we focus on acoustic IFE [Su2025]. Related case of a chiral strain induced time dependent axial electric field $E_\omega^5$ producing magnetization, the so-called "5th IFE" was also discussed [Lia2021]:

$$M \sim E_\omega^5 \times E_\omega^{5\,*} \qquad\qquad \text{A.3.1}$$

Emergent Gauge Fields in Dirac Insulators: Dirac insulators such as hexagonal boron nitride (hBN) or transition-metal dichalcogenides host low-energy quasiparticles that behave as massive Dirac fermions [Xia2007]. Importantly, these systems often preserve time-reversal symmetry (TRS) but break inversion symmetry, enabling the definition of valley-contrasting Chern numbers. When the crystal lattice is deformed by acoustic waves lattice deformations generate a valley-dependent vector potential $\mathbf{a} = g\left(u_{yy} - u_{xx},\ 2u_{xy}\right)$, acting as an effective gauge field [Voz2010].This is reminiscent of an electromagnetic vector potential but is purely geometrical in origin, stemming from strain-induced modulation of hopping amplitudes. Crucially, this field is odd under valley exchange and thus respects TRS while enabling topological valley currents.

Due to TRS, the response of the Dirac insulator to both strain and the physical electromagnetic field **A** is governed by a cross Chern-Simons action [Su2025]:

$$S_{CS} = \frac{sgn(m)}{2\pi} \int d^3x \varepsilon^{\mu\nu\lambda} A_\mu \partial_\nu a_\lambda. \qquad\qquad A.3.2$$

The diagonal Chern-Simons terms—those involving only **A** or only the emergent gauge field **a** due to strain —vanish as a consequence of TRS. This results in the electric charge and current response to strain field: , and $\rho = -\left(\frac{2e^2}{h}\right)\mathrm{sgn}(m)\mathbf{B}_s$, where the effective pseudo-electric and pseudo-magnetic fields are defined as $\mathbf{E}_s = -\partial_t \mathbf{a}$ and $\mathbf{B}_s = (\nabla \times \mathbf{a}) \cdot \hat{z}$. Here, $m$ is the Dirac mass, and $\mathrm{sgn}(m)$ determines the valley Chern number that encodes the material's topological character.

Static Magnetization via Circularly Polarized Phonons: The phonon inverse Faraday effect is realized when a circularly polarized surface acoustic wave propagates through a periodically rippled Dirac insulator (e.g., hBN) placed on a piezoelectric substrate, as sketched in Fig. 10. The acoustic wave's rotation drives the strain-induced charge density into orbital motion, generating a dc magnetization: , where $\rho(\mathbf{r})$ is the charge density induced by pseudo magnetic field generated by strain, and ***u*** is the displacement field from the surface acoustic wave.

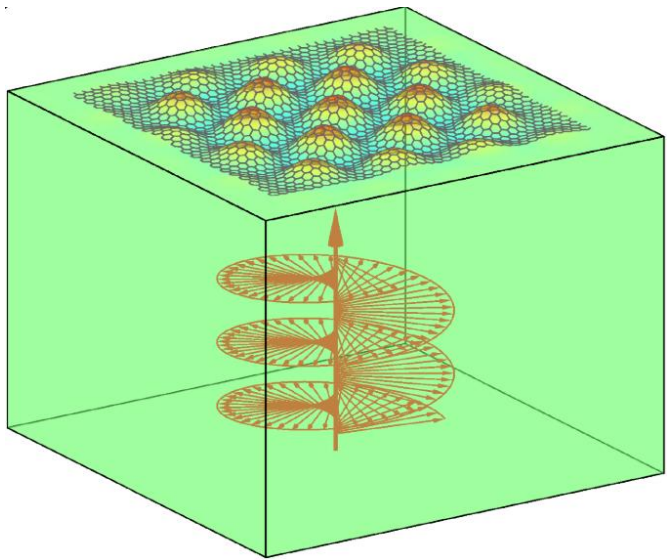

**Fig. 10:** schematic setup for the acoustic IFE. A periodically rippled 2D Dirac material is placed on a piezoelectric substrate in which a circularly polarized acoustic wave propagates in the out-of-plane direction. This figure is adapted from Ref. [Su2025].

Unlike conventional IFE, the acoustic IFE in Dirac insulators scales linearly with acoustic frequency and is independent of quasiparticle lifetime, highlighting its topological protection. [Su2025] This represents a form of topological IFE, where the direction of induced magnetization depends explicitly on the sign of the Dirac mass for each valley. Since the magnetization scales with sgn($m$), this means the effect is controlled by the topological index—the valley Chern number—rather than microscopic details.

The acoustic IFE represents a compelling form of acoustic QP, where quantized vibrational modes induce non-thermal, topologically robust, and spatially structured magnetization. Since this effect does not require doping or spin-orbit coupling and works in insulating, room-temperature platforms, it is promising for future phononic logic, valleytronic control and topological device applications.

### A4. Quantum Printing in Transverse Quantum Fluids (QHE Systems)

While discussing the IFE in metals and superconductors, we implicitly assumed that the dominant response to the light's electric field is a longitudinal current. However, the presence of a significant transverse response gives rise to qualitatively new effects, which we now discuss.

A famous example of a transverse fluid is the quantum Hall (QH) effect. At the level of linear response, it has led to many novel insights into the role of topology and broken symmetries in electronic transport [Von2020, Has2010]. At the center of the QH plateaus, the DC linear response is purely transverse, and the Hall conductivity $\sigma_{\omega=0}^H$ is quantized by the Chern number invariant of the electronic wavefunctions [Tho82]. The AC response includes also a purely reactive longitudinal component $\sigma_\omega^L$, so that total transport is still perfectly dissipationless. In intrinsic Chern insulators, the quantum Hall effect can be realized in the absence of a magnetic field because the material itself breaks time-reversal symmetry [Tok2019, Ju2024]. The nonlinear Hall effect demonstrates transverse nonlinear response [Du2021].

These features also appear in quantum printing, which is enriched by transverse flow. Because in the purely transverse case the current remains orthogonal to the applied field, circularly polarized light still drives rotating currents (see Fig. 11(a)). This leads to a transverse contribution to the IFE, where the static magnetization is given by $M_0 = i\frac{|\sigma_\omega^H|^2}{2e\omega\rho_0}E_\omega \times E_\omega^*$. In the QH fluid, this term dominates the IFE and can induce magnetizations of the order of one Bohr magneton per carrier in graphene and transition-metal dichalcogenides (TMDs) under illumination by THz light [Car2025].

A qualitatively distinct effect is the generation of magnetization under linearly polarized light, known as the inverse Cotton-Mouton effect (ICME). While it is typically attributed to the light-induced spin polarization in magnetic systems [Kir2010], an orbital analog exists in transverse fluids, where

$$M_0 = \frac{Im(\sigma_\omega^L \sigma_\omega^{H\cdot})}{e\omega\rho_0} |E_\omega|^2 \qquad\qquad\qquad A.4.1$$

is the magnitude of the induced out-of-plane magnetization[Car2025]. Note that this expression is non-zero only if both longitudinal and transverse current responses are present and are out-of-phase. This means that the optical response of the material is chiral, as indeed is the case in the QH example where $\sigma_\omega^H$ is real and $\sigma_\omega^L$ is imaginary. To understand the connection to chirality, note that a linearly polarized field $E_\omega = E_0 \cos(\omega t)\, \hat{e}_x$ generates a rotating current $J_\omega = E_0\big(Im[\sigma_\omega^L]\sin(\omega t)\, e_x + Re[\sigma_\omega^H]\cos(\omega t)\, e_y\big)$ , with chirality determined by the sign of $Im[\sigma_\omega^L \sigma_\omega^{H\cdot}]$, in the same way as the sign of the ICME in the nonlinear response (A3.1) (see Fig. 11b))). In comparison with quantum printing, where structured light can imprint symmetry-breaking onto matter, the ICME can be understood as a quantum version of photographic development: unstructured, symmetric light acts as a developer which makes the underlying symmetry-breaking of the material visible.

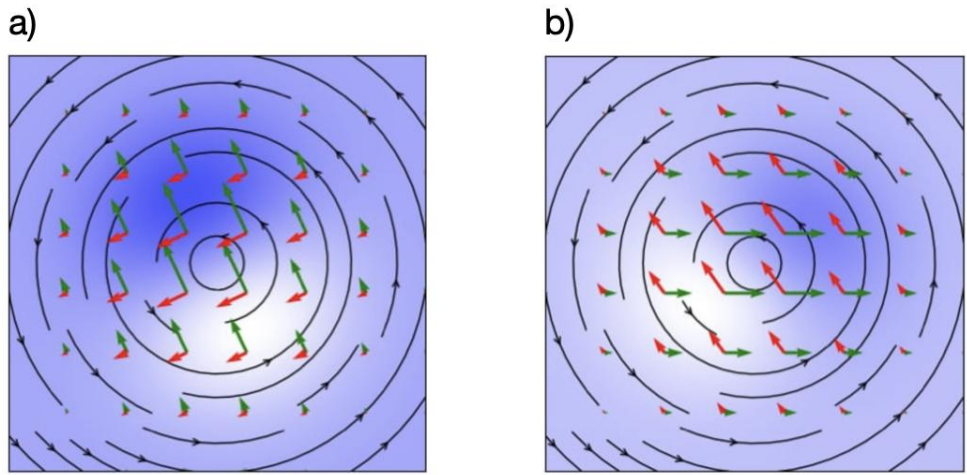

**Fig. 11:** transverse component of (a) the IFE and (b) the ICME. The green arrows represent the oscillating electric field, red arrows represent the oscillating velocity of the fluid, and the color gradient represents the oscillating background density. The superposition of the oscillating velocity and density leads to a rectified static vortical current (black stream plot), corresponding to a static orbital magnetization. See supplementary material for the animations. (Movie S3 in Supplemental Material [SM])

### A.5. Quantum Turbulence.

QP can lead to novel far from equilibrium collective behavior in a controllable fashion. One exciting frontier involves accessing turbulent behavior – a hydrodynamic regime in which nonlinearity in the collective dynamics of the superconductor dominates over viscous damping and induces energy transfers across multiple length scales. An outstanding question to be explored is whether intrinsically quantum dynamics via Heisenberg's equation of motion with non-commuting operators results in new patterns of turbulent energy transfers that are absent in classical fluid dynamics.

Turbulence of quantum fluids has mostly been discussed in the context of neutral superfluid Helium-4 [Bar2014,SKR2021]. We propose to use the beam to induce turbulent regimes in charged superconducting fluids. The gauge potential of light can stir the electron liquid and induce strongly disordered phase regimes. Moreover, since the wave vector associated with light can far exceed the microscopic scales associated with the superconductor (e.g. the coherence length), the stirring can occur at large distances, in a similar way to stirring of classical fluids in the turbulent regime.

We illustrate the concept of QP induction of quantum turbulence using the example of light-induced vorticity and phase fluctuations in the superconducting film. Consider the case of superconductor subject to light, as shown in Fig 12. Our preliminary results indicate that the long enough exposure to the QP beam can lead to turbulent regime in the superconducting case with the phase correlations $C_{JJ}(r, r') = \langle J_s(r) J_s(r') \rangle$ decaying with distance in a power law.

QP methods enables to ask whether turbulent hydrodynamics can arise in relaxational microscopic dynamics as in time-dependent Ginzburg Landau theory. In such cases, non-linearity and hydrodynamic behavior emerges in a coarse-grained theory while absent in the microscopic theory. Such emergent behavior is certainly not unprecedented in condensed matter systems; a well-known example is the emergent relativistic invariance in quantum spin systems tuned to critical points.

The detailed analysis and possible origins and nature of turbulent cascades in quantum charged fluids would need to be tested. Light induced flows of charged fluids and the induced quantum turbulence would present an interesting application of the QP ideas we discuss here.

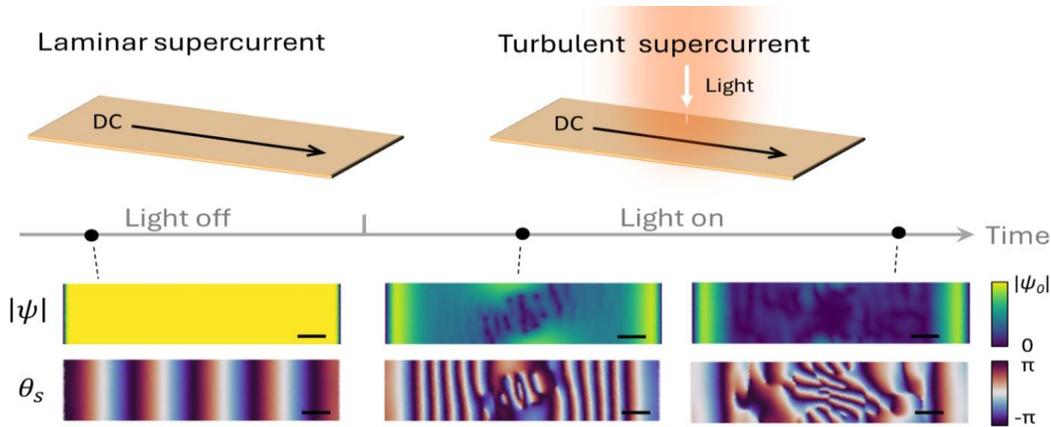

**Fig 12:** Optical driven laminar to turbulent change: left panel shows laminar supercurrent before light on , right panel shows turbulent flow. The simulation shows the amplitude $|\psi|$ and phase $\theta_s$ of superconducting order parameters obtained from the numerical simulation of the time-dependent Ginzburg-Landau model. The transition is vividly seen in the phase profile, lowest panel. The scalar bars of data represent 2 superconducting coherent length.

## B. Quantum printing in neutral coherent fluids

So far, we have discussed QP through the interaction of the electric field of light with charged matter. However, QP is also possible in systems that are overall charge neutral, where light interacts with electric and magnetic dipoles, as well as higher-order multipoles arising from charge distributions and spin-induced magnetic moments.

### B1.  Quantum printing in neutral coherent fluids

An example of such a charge-neutral system is the spatially separated particle-hole exciton fluid in transition metal dichalcogenide (TMD) bilayers. Spatially separated excitons are formed by electrons in one layer bound by Coulomb interaction to holes in the other layer, separated by a distance $d \sim 10\ \text{Å}$ [Jia2021]. In these systems, the large electric dipole moment of excitons can dominate the optical response even at room temperature [Mue2018].

Hence, we can ask how to realize QP in a dipolar fluid where there are only particle-hole dipoles and no free charges. One can show that the rectified magnetization generated by stirring the local electric dipole moments is

$$M_0 = n_0 Re(v_\omega \times P_{-\omega}),$$

B1.1

where $P_\omega$ is the particle-hole dipole vector at frequency ω, $v_\omega$ is the midpoint velocity of the dipole, and $n_0$ the density of dipoles. This expression corresponds to the IFE for dipolar fluids and illustrates the QP of magnetization through the induced rotation of electric dipole moments. The contribution can also be understood in terms of the induced currents on each layer [Syl2025]. Because each particle-hole dipole is charge neutral, one might expect that the opposite orbital magnetic moments of electrons and holes in the two layers yield zero net magnetization. However, in bilayer TMDs, electrons and holes can possess different effective quasiparticle masses and each layer can have a distinct coupling strength to the electromagnetic field, both of which contribute to the rectified magnetization. Thus, the contributions in the two layers may not cancel, but instead leads to a net significant IFE. Fig.13 illustrates the proposed mechanism. Therefore, looking ahead, the IFE and the associated induced magnetization may provide a new way to probe exciton formation in TMD bilayers.

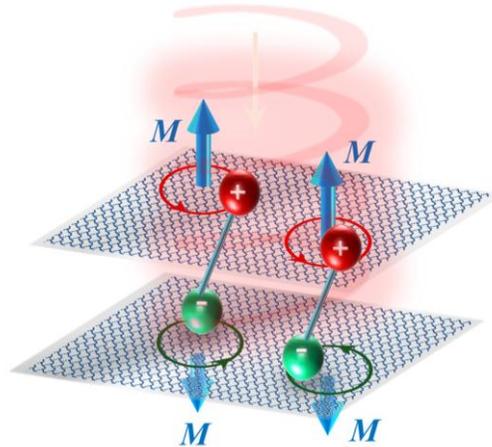

**Fig 13:** IFE in bilayer TMD with excitons. Circularly polarized light generates rotation of electric dipole moments in the bilayer. This leads to opposite orbital magnetic moments for electrons in the top layer and holes in the bottom layer. However, the net magnetic moment remains nonzero due to the asymmetry between electrons and holes in their effective masses and coupling strengths to the electromagnetic field.

## B2. Quantum printing in magnetic materials – skyrmion printing.

There are few established approaches that lead to skyrmion textures in magnets: i) use of equilibrium chiral magnets with bulk or interfacial Dzyaloshinskii–Moriya interaction that stabilise skyrmion lattices under modest magnetic fields and in a narrow temperature range, as first observed in MnSi and later in Fe/PdFe bilayers on Ir(111) [Müh2009,Hei2011], ii) use of spin-transfer and spin-orbit torques from current pulses nucleate isolated skyrmions in multilayers or antiferromagnetic heterostructures [Jia2015]; iii) and recently proposed use of light-induced heat pulses that trigger local demagnetization, and subsequent realignment of spin textures into a skyrmions in chiral magnets. This route has been now thoroughly tested to be used in a pump–probe x-ray imaging of single-defect dynamics [Ber2018]. All these techniques ultimately rely on breaking inversion symmetry through materials chemistry, interfaces or thermal gradients; they would struggle in centrosymmetric lattices and offer limited spatial selectivity. We see the new route to skyrmion creation using structured LG beams.

An exciting new opportunity would be the use of QP through the interaction of structured light with uncharged matter to print non-collinear texture to magnetic materials. Localized spins are charge neutral object. In such systems, it is the magnetic field of structured light that would couple to the localized spins and allow for QP. The time dependent magnetic field produces a torque on the localized spins and imprints the non-collinear spin textures such as vortices and skyrmions. As QP relies on the spatial and temporal structure of the magnetic field of the beam, it is not necessary to have the broken symmetries and/or non-collinear generating interactions like anisotropic, and Dzyaloshinksii-Morya interactions. Our preliminary results show that one can induce local skyrmion density in a collinear ferromagnet using a LG beam, Fig. 14.

Our results indicate Laguerre–Gaussian (LG) pulse carrying orbital angular momentum (OAM) index l generates an in-plane magnetic field proportional to $e^{il\varphi}$, transferring its winding number directly to the spin texture. Using Landau–Lifshitz–Gilbert (LLG) equation

$$\partial \boldsymbol{M}/\partial t = -\gamma \boldsymbol{M} \times \boldsymbol{H}_{eff} + \alpha \boldsymbol{M} \times \partial \boldsymbol{M}/\partial t \qquad \text{B.2.1}$$

where $\boldsymbol{M}$ is the unit vector representing the direction of magnetization, γ the gyromagnetic ratio, α the Gilbert damping, and

$\boldsymbol{H}_{eff}$ is the effective field including light induced magnetic field. Previous LLG spin dynamics simulations show that a light-induced heat pulse nucleates |l| skyrmions whose core polarity is dictated by the sign of $l$, while $l = -1$ relaxes into a skyrmionium ring [Fuj2017]. The conversion occurs on much faster time scale than thermal-quench or current-driven induced skyrmions. By properly engineering the beam spot same material can host distinct defect families selected solely by the optical phase mask. Extending above concept beyond chiral magnets with Dzyaloshinskii–Morya interaction (DMI), our recent micromagnetic simulations demonstrate that, a LG beam with OAM index $l = 1$ and spin angular momentum (SAM) index $s = 1$, directly applied to a collinear magnet without DM does imprints a skyrmion

[Liu2025]. Fig. 14 visualizes the resulting magnetization texture and its topological-charge density

$$q(r,t) \;=\; \tfrac{1}{4\pi}\boldsymbol{M}(r,\;t)\cdot\Big(\partial_x\boldsymbol{M}(r,t)\times\partial_y\boldsymbol{M}(r,t)\Big) \qquad\qquad \text{B.2.2}$$

Looking ahead, QP offers a materials-agnostic knob to integrate topological magnetic spin textures with existing magnonic systems. A skyrmion texture has been demonstrated. during the micromagnetic simulation of spin dynamics through the interaction between l = 1, s =1 mode LG beam and collinear magnetic spin system, in which the Dzyaloshinskii–Moriya interaction is turned off. An illustration of Néel type skyrmion is shown in the bottom right corner. Finally, we remark that QP can be used to reverse magnetization thus demonstrating its ability to use QP to "erase" the magnetization and defects.

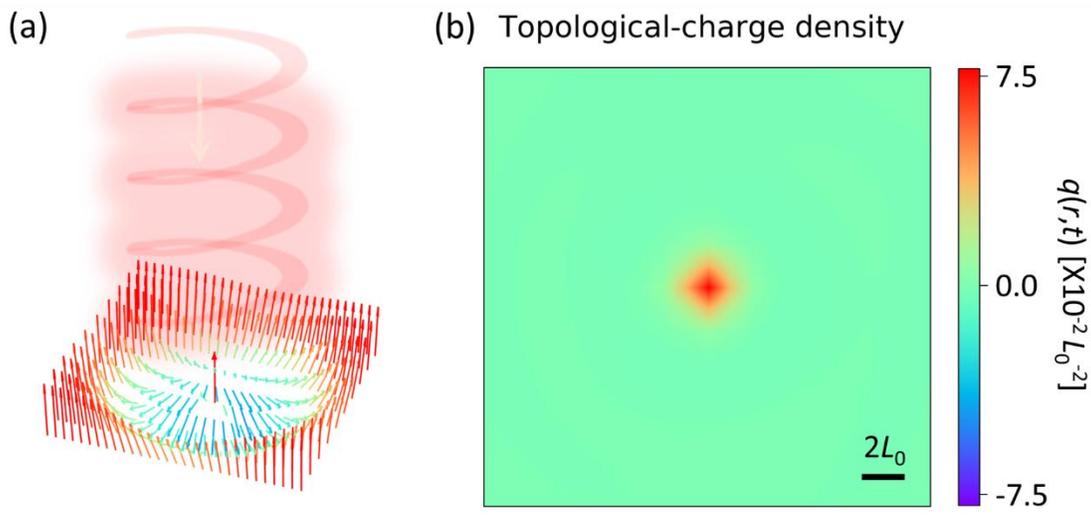

**Fig. 14:** The $s$=1, $l$=1, $p$=0 mode LG beam induced skyrmion is captured in the LLG spin dynamics simulation. The scalar bar represents 2 lattice sites ($L_0$). The skyrmion texture is shown in panel (a) and the corresponding topological skyrmion density $q$(r,t) of the spin texture is shown at some point in the light cycle in (b). Dynamics of the spins shows also its ability of a structured light to reverse magnetization of spins. (Movie S4 in Supplemental Material [SM])

**Outlook and Future directions.**

This overview outlined the ideas and motivation behind the new approach to manipulate quantum matter- the quantum printing. Although rooted in history and long practiced in AMO, the concept of quantum printing is new and rapidly evolving. We overview the current progress in using QP to create new states and excitations in coherent charged fluids like metals, superconductors and transverse Hall fluids and discuss the cases of neutral magnetic and excitonic fluids. We also aim to lay out some future directions of applications and to provide timely discussion for the rapidly growing field.

QP of the new states and new excitations can be accomplished with structured light and with structured strain. The strain component of QP is much less developed and we expect new exciting results here. Structured strain and sound modes would enable this progress.

We believe the concept of QP can provide an importance guidance in developing further directions of research of quantum states. We see directions of research on QP developing.:

**Materials Design**: using platforms like 2D materials, moiré systems for high-efficiency quantum print/rectification. For example, we expect quantum printing with LG beams also can induce Higgs modes in superconductors [Kan2025]. Another interesting topic one might explore is the quantum printing of excitonic orders like interlayer exciton states discussed and the Rydberg excitons.

**Strain and Acoustic IFE and printing**. We see equally exciting opportunities in QP using the structured sound and strain. Experimental challenges that one would need to address include: For realistic parameters—acoustic frequencies in the GHz range, ripple heights of a few nanometers, and acoustic displacements of ~10 nm—the acoustic IFE produces magnetization on the order of 1 nuclear magneton per unit cell, a level comparable to magnetization from optical phonons in ionic materials [Jur2017]. The geometry of the substrate or strain pattern would control the periodicity and symmetry of the induced magnetization. A simpler uniform magnetization can be induced via triaxial stretching.

**Light induced topological magnonics**. Our results support the notion of skyrmion printing using structured light. Key open questions in this approach include how beam fluence and duration set the threshold for defect proliferation, how the method scales in antiferromagnets where net Zeeman coupling vanishes, and whether synchronized multi-beam interference can seed skyrmion crystals on demand. Answering these questions will move optical phase engineering from a proof-of-principle towards quantum printing memory technology.

**Quantum Control**: For example - IFE can be used as the quantum state initiation for quantum computation: the light via IFE can imprint a fully polarized state of semiconducting and q dot states to start with $|\Psi\rangle = \otimes_1^N |\uparrow\rangle_i$.

**Quantum Turbulence**: light induced phase changes in coherent fluid can lead to a formation of turbulent states in the quantum regime. The light induced turbulence would naturally occur in electron fluids like superconductors and quantum hall fluids.

**Cavity Quantum Materials**: The unexplored connections of the quantum printing ideas to the new field of cavity quantum materials remains to be explored and discussed.

**Multimode and in-operando device applications**. Another important direction to be pursued is quantum printing in multi-mode implementations. For instance we envision the scaling of quantum printing to hybrid systems where the light induced states can be controlled and probed using the electronic or spintronic readouts. These applications open new perspectives for in-operando applications.

**Light- Reconfigurable circuits.** Another potential application is the use of QP to print and change the circuits. Ability to print also implies the ability to erase and change the patterns created in quantum material. Hence one can use the light and strain to change the pattern of electronic and magnetic states. Hence we foresee potential to use QP to create reconfigurable states and circuits. One of the simplest tests of this approach would be ability to erase vorticity in superconductors.

**QP with entangled states**. QP with entangled states could lead to printing with high temporal resolution, while minimizing transition barriers, and enhancing selectivity for the desired state. Inspiration can be drawn from molecular femtochemistry, where early studies demonstrated how sequential femtosecond pulses offers a knob to effectively control the course of a chemical reaction.[Zew2000]. In condensed matter systems, where this concept remains in its early stages of exploration, a femtosecond optical pulse is used to excite the electronic structure, transiently modulating the ground state potential energy surface. This can lead to changes in the relative stability of metastable phases—corresponding to local minima in the energy landscape—and in the heights of activation barriers, thereby enabling access to these phases via excitation with a second pulse, Fig 2. Current state-of-the-art has successfully demonstrated the connection between a 2nd optical excitation in-phase with a coherently excited phonon mode and increased switching efficiency, indicative of a nonthermal collective transition pathway.[Hor2020b,Mak2023].

Applications of QP might emerge that we have not thought of. Ultimately the impact of the QP as a concept could be in guiding the design and controlling states of quantum matter with "on-demand" functionalities.

*Acknowledgements*: We are grateful to our colleagues and our research groups for critique and advise in developing these ideas. This work was supported by US DOE Office of Basic Energy Sciences under Award No. DE-SC-0025580 (AB, PW, TY). Work of GA, GC, YL, was supported by European Union Seventh Framework ERC-2018-SYG 810451 HERO, the Knut and Alice Wallenberg Foundation KAW 2019.0068. Work of SZL was carried out under the auspices of the U.S. DOE NNSA under contract No. 89233218CNA000001 through the LDRD Program, and was performed, in part, at the Center for Integrated Nanotechnologies, an Office of Science User Facility operated for the U.S. DOE Office of Science, under user proposals #2018BU0010 and #2018BU0083. Work of ES was supported from the Norwegian Research Council through Grant No. 262633, "Center of Excellence on Quantum Spintronics". Work of JW was conducted with support from Vetenskapsrådet under contract 2021-04379.

*Software*: For the results of simulations of light imprinting in superconductor we developed the software available at Github open source: LG-TDGL provides a platform to explore the interaction between structured light and superconducting order parameters, enabling the realization of the quantum printing effect. We showcase an example of the dynamics of superconducting vortices induced by Laguerre-Gaussian beam and demonstrate the results for the light-imprinted superflow. (Link: https://github.com/TienTienYeh/lg-tdgl)